\documentclass[12pt]{article}
\usepackage{amsmath,amsfonts}
\makeatletter \@addtoreset{equation}{section}

\usepackage{cite}
\usepackage{bm}
\usepackage{dcolumn}
\makeatother
\def\one{{\hbox{ 1\kern-.8mm l}}}
\newcommand{\Dslash}{\not{\hbox{\kern-4pt $D$}}}
\newcommand{\pdslash}{\not{\hbox{\kern-2pt $\partial$}}}
\newcommand{\be}{\begin{equation}}
\newcommand{\bea}{\begin{eqnarray}}
\newcommand{\eea}{\end{eqnarray}}
\newcommand{\ba}{\begin{array}}
\newcommand{\ea}{\end{array}}
\newcommand{\ee}{\end{equation}}
\newcommand{\bzeta}{\mbox{Z\hspace{-0.8em}Z}}
\begin{document}
\begin{titlepage}
\vspace*{1mm}%
\hfill%
\vbox{
    \halign{#\hfil        \cr
           IPM/P-2009/008 \cr
                     } 
      }  
\vspace*{15mm}%
\begin{center}

{{\Large {\bf  Asymptotic symmetry of geometries with Schr\"odinger isometry }}}

\vspace*{15mm}
\vspace*{1mm}
{Mohsen Alishahiha$^a$, Reza Fareghbal$^a$, Amir E. Mosaffa$^{b,a}$ and Shahin Rouhani$^{b}$}

 \vspace*{1cm}
{\it ${}^a$ School of physics, Institute for Research in Fundamental Sciences (IPM)\\
P.O. Box 19395-5531, Tehran, Iran \\ }

\vspace*{.4cm}

{\it ${}^b$ Department of Physics, Sharif University of Technology \\
P.O. Box 11365-9161, Tehran, Iran}

\vspace*{.4cm}
{  alishah, mosaffa, rouhani@ipm.ir, fareghbal@theory.ipm.ac.ir} 

\vspace*{2cm}
\end{center}

\begin{abstract}
We show that the asymptotic symmetry algebra of geometries with Schr\"odinger isometry in any 
dimension is an infinite 
dimensional algebra containing one copy of Virasoro algebra. It is compatible with the fact that 
the corresponding geometries are dual to non-relativistic CFTs whose symmetry algebra is the Schr\"odinger 
algebra  which
admits an extension to an infinite dimensional symmetry algebra containing a Virasoro subalgebra. 
\end{abstract}

\end{titlepage}

\section{Introduction }

AdS/CFT correspondence \cite{Maldacena:1997re} has provided us with a powerful framework
to study strongly coupled conformal field theories. This is done by making use of  weakly coupled gravities
on backgrounds containing an AdS part. According to the AdS/CFT duality
there is a one to one correspondence between objects on the gravity side and those in the dual conformal 
field theory. In particular it is known that the symmetries of the conformal field theory can be
geometrically realized on the gravity side as the isometries of the metric. Indeed it is well known that 
the asymptotic symmetry of an $AdS_{d+1}$ geometry is the conformal group in $d$ dimensional 
spacetime \cite{Brown:1986nw}.

For later use, following \cite{Brown:1986nw}, it is instructive to review how to find the asymptotic symmetry 
of a geometry which is asymptotically AdS. Consider a metric $g_{\mu\nu}={\bar g}_{\mu\nu}+h_{\mu\nu}$ where ${\bar g}_{\mu\nu}$ is the metric of an 
$AdS_{d+1}$ space given by
\be
ds^2={\bar g}_{m n}dx^m dx^n=\frac{-dt^2+dx_i^2+dz^2}{z^2},\;\;\;\;\;\;i=1,\cdots,d-1.
\ee
and $h_{\mu\nu}$ is deviation of the metric consistent with the particular solution.
To have a well defined asymptotically AdS solution $h_{\mu\nu}$ should have a proper falloff as
we approach the boundary at $z\rightarrow 0$. We consider
the boundary conditions as follows \cite{Brown:1986nw}
\be\label{bAdS}
h_{\mu\nu}={\cal O}(1),\;\;\;\;\;\;h_{z\mu}={\cal O}(z),\;\;\;\;\;\;h_{zz}={\cal O}(1),
\ee
where $\mu,\nu=0,1,\cdots d-1$. Now the aim is to find the asymptotic Killing vectors $\xi=\xi^m\partial_m$ which
preserve the above boundary conditions. It is straightforward to solve the asymptotic Killing equations,
${\cal L}_\xi{\bar g}_{mn}=h_{\mu\nu}$, to find \cite{Gubser}
\bea
\xi^\mu &=&\epsilon^\mu(t,x_i)-
\frac{z^2}{d}\eta^{\mu\nu}\partial_\nu \left(\sum_\rho\partial_\rho \epsilon^\rho(t,x_i)\right)
+{\cal O}(z^4),\cr
\xi^z&=&\frac{z}{d}\sum_\rho \partial_\rho\epsilon^\rho(t,x_i)+{\cal O}(z^3),
\eea
where $\eta^{\mu\nu}$ is the metric of the flat Minkowski space and $\epsilon^\mu$ is a 
function of $t$ and $x_i$ satisfying
\be
\partial_\mu\epsilon_\nu(t,x_i)+\partial_\nu\epsilon_\mu(t,x_i)=\frac{2}{d}\sum_\rho\partial_\rho\epsilon_\rho.
\ee
We note that this is the equation we get for the parameter of the conformal transformation in $d$ dimensions. When
$d>2$ this leads to $SO(2,d)$ group while for $d=2$ the conformal group generates two copies of the 
Virasoro algebra.

The aim of this article is to extend the above considerations for geometries with Schr\"odinger isometry.
In fact it has recently been proposed in \cite{{Son:2008ye},{Balasubramanian:2008dm}} 
that gravity on a background with Schr\"odinger isometry is dual
to a non-relativistic CFT. 

An important point we would like to mention is  
that the $d-1$ dimensional Schr\"odinger algebra which generates the symmetry of 
the $d$ dimensional non-relativistic CFTs admits an extension to an infinite dimensional symmetry algebra 
containing a Virasoro  subalgebra. This has to be compared with the relativistic CFTs whose symmetry
algebra is finite in all dimensions except for $d=2$, where we get two copies of Virasoro algebra.
Therefore if we would like to establish the non-relativistic AdS/CFT correspondence one should be able
to show that the asymptotic symmetry of geometries with Schr\"odinger isometry is an 
infinite dimensional algebra containing one copy of the Virasoro algebra in arbitrary dimensions.
In this article we show how this can happen.

This paper is organized as follows. In the next section we briefly review the $d-1$ dimensional Schr\"odinger
algebra and its infinite dimensional extension. In section three we study the asymptotic 
symmetry of the geometries with Schr\"odinger isometry where we show that an infinite dimensional 
symmetry can be realized. To be precise in this section we will obtain the algebra of the 
asymptotic Killing vectors. We note, however, that the conserved charges generating the asymptotic 
symmetry in general have the same algebra up to central extension. The last section is devoted to discussions.

\section{Non-relativistic CFTs and Schr\"odinger algebra}

In this section we  review the Schr\"odinger algebra and its infinite dimensional
extension in arbitrary dimensions. 
The generators of the Schr\"odinger algebra are spatial translations $P_i$, rotations $J_{ij}$, time translation $H$, 
Galilean boosts $K_i$, dilation $D$ and special conformal transformation $C$. 
The algebra of $Sch_{d-1}$ with a central extension given by the number operator $N$ may be
written as \footnote{This algebra can be obtained from the relativistic conformal 
algebra in $d+1$ dimensions. In other words the 
Schr\"odinger group may be 
thought of as a subgroup of $SO(2,d+1)$ with fixed momentum along the null direction 
(see for example \cite{{Burdet:1977qw},{Duval:1984cj},{Henkel:2003pu},{D3}}).}
\bea
&&[J_{ij},P_k]=-(\delta_{ik}P_j-\delta_{jk}P_i),\;\;\;\;\;\;\;\;\;\;\;\;\;\;\;\;\;\;
[J_{ij},K_k]=-(\delta_{ik}K_j-\delta_{jk}K_i),\cr
&&[J_{ij},J_{kl}]=-\delta_{ik}J_{jl}+{\rm perms},\;\;\;\;\;\;\;\;\;\;\;\;\;\;\;\;\;\;[P_i,K_j]=-N\delta_{ij}\cr
&&[D,P_i]=\frac{1}{2}P_i,\;\;\;\;\;\;\;\;[D,K_i]=-K_i,\;\;\;\;\;\;\;\;[D,H]=H,\cr
&&[C,P_i]=K_i,\;\;\;\;\;\;\;\;\;\;\;[D,C]=-C,\;\;\;\;\;\;\;\;\;\;[C,H]=2D.
\eea
It is worth noting that the generators $D,H$ and $C$ form an $SL(2,R)$ algebra. As we will see this is 
exactly the reason why the Schr\"odinger algebra has an infinite dimensional extension.
 On the other hand 
since $[D,N]=0$ one may diagonalize them simultaneously leading to the fact that representations of 
the Schr\"odinger algebra may be labeled by two numbers; conformal dimension $\Delta$ and a number $M$ which are
the eigenvalues of $D$ and $N$, respectively.

The generators of the Schr\"odinger algebra can be thought of as vector fields defined on 
$d$ dimensional spacetime with the following representation
\bea
&&J_{ij}=-(x_i\partial_j-x_i\partial_i),\;\;\;\;\;\;\;\;P_i=-\partial_i,\;\;\;\;\;\;\;\;\;
H=-\partial_t,\;\;\;\;\;\;\;\;N=-M\cr
&&K_i=-(t\partial_i+x_i M),\;\;\;\;\;D=-(t\partial_t+\frac{1}{2}x_i\partial_i),\;\;\;\;\;
C=-(t^2\partial_t+tx_i\partial_i+\frac{1}{2} x^2 M).
\eea
where $x^2=x_ix_i$. Following \cite{Henkel} one can define the generators of the corresponding infinite dimensional algebra in $d-1$ dimensions as follows
\bea\label{gen}
L_n&=&-t^{n+1}\partial_t-\frac{n+1}{2}t^{n} x_i\partial_i-\frac{n(n+1)}{4}t^{n-1} x^2 M,\cr
Q_{i{\hat n}}&=&-t^{{\hat n}+1/2}\partial_i-({\hat n}+\frac{1}{2})t^{{\hat n}-1/2}x_iM,\;\;\;\;\;T_n=-t^n M.
\eea
Here $n\in \bzeta$ and ${\hat n}\in \bzeta+\frac{1}{2}$. It is straightforward to see
that the above generators satisfy the following commutation relations
\bea\label{algebra}
&&[L_n,L_m]=(n-m)L_{n+m},\;\;\;\;\;\;\;\;\;\;\;\;\;
[Q_{i{\hat n}},Q_{j{\hat m}}]=({\hat n}-{\hat m})\delta_{ij}T_{{\hat n}+{\hat m}},\cr
&&[L_n,Q_{i{\hat m}}]=(\frac{n}{2}-{\hat m})Q_{i(n+{\hat m})},
\;\;\;\;\;\;\;\;[L_n,T_m]=-mT_{n+m}.
\eea
Note that due to non-trivial contribution of number operator to the Galilean boost the Schr\"odinger 
algebra does not allow an infinite dimensional extension for rotations $J_{ij}$. Thus we find
\be
[J_{ij},Q_{k {\hat n}}]=-(\delta_{ik}Q_{j\hat{n}}-\delta_{jk}Q_{i\hat{n}}),\;\;\;\;\;\;[L_n,J_{ij}]=0.
\ee
This may be compared with the Galilean conformal algebra recently studied in \cite{Bagchi:2009my}
where it was shown that the algebra allows an infinite dimensional extension for rotations too.

In the next section we show that the above algebra is indeed asymptotic symmetry algebra of the 
geometries with Schr\"odinger isometry.

\section{Asymptotic symmetry}

Following the relativistic AdS/CFT correspondence one expects that if there exist gravity duals to  
non-relativistic CFTs, the isometry of the relevant geometry must be the  Schr\"odinger group. In fact
such gravity duals exist, for example consider the metric  
\be\label{sonic}
ds^2=-\frac{dt^2}{z^4}+\frac{-dtd\eta+dx_i^2+dz^2}{z^2},\;\;\;\;\;i=1,\cdots,d-1,
\ee
which could be thought of as a solution of a $d+2$ dimensional gravity coupled to massive
gauge field \cite{Son:2008ye,Balasubramanian:2008dm}. It is easy to see that the isometry of the  
above metric has $d-1$ dimensional Schr\"odinger group.

This metric has been proposed to provide a gravity description for non-relativistic CFTs.
Such non-relativistic CFTs admit a Schr\"odinger algebra. On the other hand since 
the Schr\"odinger algebra admits an infinite dimensional extension one may wonder if the extended algebra can be
seen from the gravity description too. The situation is very similar to the $AdS_3$ case where it is known that its
asymptotic symmetry algebra contains  two copies of Virasoro algebra in agreement with the symmetry of 2D CFT.

In this section we study the asymptotic symmetry of the metric \eqref{sonic} to see how 
the infinite dimensional symmetry algebra arises. It has to be compared with the AdS space where 
the symmetry is finite dimensional except for $AdS_3$ and $AdS_2$.

The non-trivial point which allows the non-relativistic CFTs to have
infinite dimensional symmetry algebra is the fact that the Schr\"odinger algebra in any dimension has an $SL(2,R)$ 
subalgebra generated by the time translation, dilation and special conformal transformation. This
factor can be extended to the Virasoro algebra and therefore to get a closed algebra the other generators
should also admit an infinite dimensional extension.

From the gravity point of view there is a direct observation to see why the metric \eqref{sonic} leads to an infinite
dimensional symmetry in any dimension while it is not the case for AdS geometry. In fact the crucial point is 
to see how the $SL(2,R)$ factor may be realized from the geometry \eqref{sonic}. To see this, it is 
useful to define a new coordinate $\rho=z^2/2$ by which the metric \eqref{sonic} reads
\be\label{AdS2}
ds^2=\frac{1}{4}\left(\frac{-dt^2+d\rho^2}{\rho^2}\right)+\frac{1}{2\rho}(-dtd\eta+dx_i^2).
\ee
It is now evident that for small $\rho$ the first factor dominates and therefore we observe that 
asymptotically the background approaches an $AdS_2$ which is the realization of the $SL(2,R)$ 
subalgebra. On the other hand it has been shown \cite{Strominger:1998yg}
that the asymptotic symmetry group of an asymptotically $AdS_2$ geometry is one copy of the Virasoro 
algebra. This is indeed the Virasoro subgroup we are 
looking for\footnote{Galilean conformal algebra in $d$ dimensions has recently been 
obtained from $d$ dimensional conformal algebra by a contraction in \cite{Bagchi:2009my}. The
authors observed that the Galilean conformal algebra admits an extension to an infinite
dimensional symmetry algebra. In the light of our discussions one may wish to 
interpret this effect as the fact that there is an $SL(2,R)$ subgroup which should also 
be geometrically realized from the gravity description. Actually this is the case. Indeed 
the corresponding gravity description may also be obtained by a contraction of $AdS_{d+1}$ geometry. Doing this 
the resultant gravity develops an $AdS_2$ factor as the base space over which we have a $d-1$ dimensional
fiber.}.

We note, however, that in our case having had an $AdS_2$ factor, though necessary,   
is not sufficient to get infinite dimensional extension of the isometry of the solution. This 
is due to the fact that the geometry is not asymptotically factorized as a direct product of $AdS_2 \times M_d$ 
with $M_d$ being  $d$-dimensional Minkowski space.  In particular we note that for arbitrary dynamical scaling 
it is always possible to bring the metric into a form with a factor of $AdS_2$, though only for the 
present case we have infinite dimensional extension of the isometry algebra\footnote{We would like 
to thank M. Edalati for pointing out the confusion of the text in the early version of the paper.}.
Therefore to find the explicit form of the  generators of the asymptotic symmetry we follow the procedure of 
\cite{Brown:1986nw} as reviewed for AdS space in the introduction. 

To proceed, motivated by the boundary conditions of the AdS space \eqref{bAdS}, we consider the 
following boundary conditions for geometries which are asymptotically equivalent to the metric
\eqref{sonic}
\be
\left(\begin{array}{llll}
h_{tt}={\cal O}(\frac{1}{z^2})&h_{t\eta}={\cal O}(1)&h_{ti}={\cal O}(z^2)&h_{tz}={\cal O}(z)\cr
h_{\eta t}=h_{t\eta}&h_{\eta\eta}={\cal O}(z^4)&h_{\eta i}={\cal O}(z^2)&h_{\eta z}={\cal O}(z)\cr
h_{it}=h_{ti}&h_{i\eta}=h_{\eta i}&h_{ij}={\cal O}(1)&h_{iz}={\cal O}(z)\cr
h_{zt}=h_{tz} & h_{z\eta}=h_{\eta z}& h_{zi}=h_{iz}&h_{zz}={\cal O}(1)
\end{array}
\right)\ .
\ee
It is straightforward to see that the following asymptotic Killing vectors preserve the above boundary condition
\bea
\xi^{(1)}&=&\bigg[\epsilon(t)+{\cal O}(z^6)\bigg]\partial_t+\bigg[\frac{z}{2}\epsilon'(t)+{\cal O}(z^3)\bigg]
\partial_z+
\bigg[\frac{x_i}{2}\epsilon'(t)+{\cal O}_i(z^4)\bigg]\partial_i\cr &&\cr
&+&\bigg[\delta(t)+\frac{z^2+x^2}{4}\epsilon''(t)+
{\cal O}(z^4)\bigg]\partial_\eta,\cr
\xi^{(2)}_i&=&{\cal O}_i(z^6)\partial_t+{\cal O}_i(z^3)\partial_z+\bigg[\beta(t)+{\cal O}(z^4)\bigg]\partial_i+\bigg[\beta'(t)x_i+{\cal O}_i(z^4)\bigg]\partial_\eta,
\eea
where $\epsilon,\beta$ and ${\delta}$ are functions of $t$ which may be thought of as independent polynomials 
of $t$. The prime denotes derivative with respect to $t$. 
To mode expand the above Killing vectors we set $\epsilon(t)=-t^{n+1},\beta(t)=-t^n$ and ${\delta}(t)=-t^n$ by which one finds
\bea
L_n&=&-t^{n+1}\partial_t-\frac{n+1}{2}t^{n}(z\partial_z+x_i\partial_i)-\frac{n(n+1)}{4}t^{n-1}(z^2+x^2)\partial_\eta,
\cr
Q_{in}&=&-(t^n\partial_i+nt^{n-1}x_i\partial_\eta),\;\;\;\;\;\;T_n=-t^n\partial_\eta,
\eea
satisfying 
\bea
&&[L_n,L_m]=(n-m)L_{n+m},\;\;\;\;\;\;\;\;\;\;\;\;\;
[Q_{i{\hat n}},Q_{j{\hat m}}]=({\hat n}-{\hat m})\delta_{ij}T_{{\hat n}+{\hat m}},\cr
&&[L_n,Q_{i{\hat m}}]=(\frac{n}{2}-{\hat m})Q_{i(n+{\hat m})},
\;\;\;\;\;\;\;\;[L_n,T_m]=-mT_{n+m},
\eea
in precise agreement with \eqref{algebra}. 
In order to compare our results with those in the 
previous section we have denoted the generators of the asymptotic Killing vectors 
by the same latter as that in \eqref{gen}. 
Note that to precisely map the resultant algebra to those in 
\eqref{algebra} one needs
to change $n\rightarrow n+\frac{1}{2}\equiv {\hat n}$ in the generators $Q_{in}$.

It is worth mentioning  that for the limit of $z\rightarrow 0$, setting $\partial_\eta=M$ the above
generators reduce to those in the previous section. This is compatible with the fact that 
the derivative $\partial_\eta$ should be identified with  the number operator in the dual non-relativistic 
CFT (see for example \cite{{Adams:2008wt},{Herzog:2008wg},{Maldacena:2008wh}}).
It is  also obvious from the metric \eqref{sonic} that it is invariant under rotations
among the spacial coordinates $x_i$'s giving rise to the rotation generators $J_{ij}$ as the
last part of the symmetry algebra.  

As a result we could  uncover the infinite dimensional Schr\"odinger algebra from the algebra
of asymptotic Killing vectors of metric \eqref{sonic}.  On the other hand we note that the
conserved charges of the corresponding asymptotic Killing vectors which generate the 
asymptotic symmetry satisfy the same algebra of the asymptotic Killing vectors up to a
possible central extension. In the next section we will give comments on the central extension 
of the resultant algebra.

\section{Discussions}

In this paper we have studied the asymptotic symmetry of geometries with Schr\"odinger isometry.
We have shown that unlike the AdS geometry where only for $AdS_3$ and $AdS_2$ we get 
infinite dimensional asymptotic symmetry, in this case the asymptotic symmetry is infinite dimensional
in any dimension.

We have noticed that the crucial point behind this interesting phenomenon is the fact that
the Schr\"odinger group in any dimension has an $SL(2,R)$ subgroup which can be 
geometrically realized from the dual gravity description. More precisely geometries with 
Schr\"ordinger isometry are asymptotically $AdS_2$. It is also known that 
the asymptotic symmetry of the $AdS_2$ space is one copy of Virasoro algebra. As a 
result the $SL(2,R)$ subalgebra can be extended to the Virasoro algebra. In fact to find the 
infinite dimensional extension of the algebra it is enough to find the Virasoro algebra. The others can be
found via the closure of the algebra.

As a byproduct, our results make it possible to compare two different proposals of the
gravity description of non-relativistic CFTs made in \cite{{Son:2008ye},{Balasubramanian:2008dm}}
and \cite{{Goldberger:2008vg},{Barbon:2008bg}}. While the former proposal is based on the 
metric \eqref{sonic}, the latter is based on gravity in the AdS geometry where the conformal 
symmetry of the boundary is broken to the non-relativistic one via a non-trivial boundary 
condition for the fields on the
bulk. As we reviewed in the introduction the asymptotic symmetry of the AdS space is a finite 
dimensional algebra which does not allow to get an infinite dimensional extension of the Schr\"odinger 
algebra. Therefore our result is in favor of the proposal based on the metric \eqref{sonic} \cite{{Son:2008ye},{Balasubramanian:2008dm}}.

Since we have a Virasoro subalgebra an immediate question arises, does this Virasoro 
algebra admit a central extension? 
In other words we would like to know if the algebra of conserved charges which generates
the above asymptotic symmetry has non-zero central charge. Actually it is known 
that there is a unique central extension to the infinite dimensional Schr\"odinger algebra \cite{Henkel} and 
indeed the central extension is the one parameterized by central charge $c$ of the Virasoro subgroup, {\it i.e.}
\be
[L_n,L_m]=(n-m)L_{n+m}+\frac{c}{12}(n^3-n)\delta_{n+m}.
\ee
On the other hand since this factor is the  common part of the Schr\"odinger algebra in any dimension,
one may expect that the non-relativistic CFTs have non-zero central charge in any dimension. Actually
it was shown \cite{Bergman:1991hf}
that the non-relativistic field theories may give rise to anomalies. 

From the gravity point of view since in any dimension the geometry develops an $AdS_2$ factor as we 
approach the boundary, following \cite{Hartman:2008dq} (see also \cite{{Alishahiha:2008tv},{Cadoni:2008mw},
{Castro:2008ms},{Alishahiha:2008rt},{Myung:2009sk}}) one may expect that the gravity on asymptotic $AdS_2$ geometry 
leads to a Virasoro algebra with non-zero central charge. Of course the argument of  \cite{Hartman:2008dq} was
based on the fact that there is a twisted energy momentum tensor due to a non-zero $U(1)$ current. Since 
the metric we have been considering may be embedded in a $d+2$ dimensional gravity 
coupled to a massive gauge filed \cite{Son:2008ye}, one may suspect that the same argument as that in 
\cite{Hartman:2008dq} may be applied in our case too leading to a non-zero central charge. It would be
interesting to understand this point better. 

To have an insight how the Virasoro algebra may get a central extension it would be instructive to 
study $d=1$ in more detail where we will also observe a new phenomenon. For $d=1$ the metric \eqref{sonic} reads
\be
ds^2=-\frac{dt^2}{z^4}+\frac{-dtd\eta+dz^2}{z^2},
\ee
with the following asymptotic symmetry algebra
\be
[L_n,L_m]=(n-m)L_{n+m},\;\;\;\;\;\;\;[L_n,T_m]=-mT_{n+m}.
\ee
The above solution has $SL(2,R)\times U(1)$ isometry which is naively extended to a Virasoro
algebra plus an affine $U(1)$ algebra. 

To find the central charge
it is useful to note that the above solution may be alternatively 
thought of as a solution of topologically massive gravity known as null $AdS_3$ solution \cite{Anninos:2008fx}. 
It was argued in \cite{Anninos:2008fx} that the CFT dual to this background at least has one copy of Virasoro algebra 
with central charge $2/G$, where $G$ is three dimensional Newton's 
constant\footnote{In our notation we set the radius of the 
metric to one.}.

In this context it has been shown \cite{Henneaux:2009pw} that the symmetry of the conserved charges may 
contain two copies of Virasoro algebra with the following central charges\footnote{We note, however,
that beside the asymptotic behavior the central charges depend on the theory as well. Therefore 
what we may find in the TMG does not necessarily apply for the model we were considering. Nevertheless
this may give an insight about the special feature of $d=1$.} 
\be
c_+=\frac{1}{G},\;\;\;\;\;\;\;c_-=\frac{2}{G},
\ee
Therefore, if correct, it means that in three dimensions we get even  more than what we bargained for; 
namely not only the central charge is non-zero but the geometry leads to two Virasoro algebras 
with non-zero central charges.

This has to be compared with higher dimensional cases where it was proved that the Schr\"odinger
algebra can have only one central term corresponding to the central extension of {\it one} copy of
 Virasoro algebra which appeared in the infinite dimensional extension of the Schr\"odinger algebra 
\cite{Henkel}. The point which might be responsible for this peculiar behavior is that for $d=1$ due
to the absence of the spacial directions it is possible to impose another falloff of the metric 
components such that to give way for bigger symmetries \cite{Henneaux:2009pw} which 
is impossible when $Q_i$'s are non-zero.

Concerning the above discussions one may conclude that similar to the relativistic CFTs for which  
$d=2$ is exceptional in the sense that conformal group becomes infinite dimensional in 
two dimensions, the non-relativistic CFT's in $d=1$ are also exceptional. While the non-relativistic 
conformal symmetry in any dimension has an infinite dimensional extension, it is only the case of 
$d=1$ where we get two copies of Virasoro algebra.
It would be extremely interesting to further explore this point.

\section*{Acknowledgments}
We would like to thank Omid Saremi for collaboration in the early stage of the project. 
We would also like to thank  Hamid Afshar, Amin Akhavan, Davod Allahbakhsi, Ali Davody and Ali Vahedi
for discussions on the different aspects of non-relativistic AdS/CFT correspondence. 
This work is supported in 
part by Iranian TWAS chapter at ISMO.



\end{document}